\documentclass[11pt]{article}

\usepackage{a4wide}
\usepackage{amsfonts, latexsym, amsmath}
\usepackage{amssymb}
\usepackage[latin1]{inputenc}
\usepackage[english]{babel}

\usepackage{graphicx}
\usepackage{url}

\newenvironment{proof}{\noindent {\it Proof.~~}\ }{\
\rule{1mm}{2mm}\medskip}

\newtheorem{thm}{Theorem}

\newcommand{\be}{\begin{eqnarray}}
\newcommand{\ee}{\end{eqnarray}}
\newcommand{\beno}{\begin{eqnarray*}}
\newcommand{\eeno}{\end{eqnarray*}}

\newcommand{\N}{\mbox{$\mathbb N$}}
\newcommand{\F}{\mbox{$\mathbb F$}}
\newcommand{\highlight}[1]{\textbf{#1}}

\begin{document}

\title{\LARGE \bf
Optimal Iris Fuzzy Sketches
}

\author{J. Bringer, H. Chabanne, G. Cohen, B. Kindarji and G. Z\'emor%
\thanks{J. Bringer and H. Chabanne are with Sagem D\'efense S\'ecurit\'e, Eragny, France.}%
\thanks{G. Cohen is with ENST, D\'epartement Informatique et R\'eseaux, Paris, France.}%
\thanks{G. Z\'emor is with Institut de Math\'ematiques de Bordeaux, Universit\'e de Bordeaux I, Bordeaux, France.}%
}

\date{}
\maketitle
\thispagestyle{empty}
\pagestyle{empty}

\begin{abstract}
Fuzzy sketches, introduced as a link between biometry and cryptography, 
are a way of handling biometric data matching as an error correction issue. 
We focus here on iris biometrics and look for the best error-correcting code 
in that respect. We show that two-dimensional iterative min-sum decoding 
leads to results near the theoretical limits. 
In particular, we experiment our techniques on the Iris Challenge Evaluation (ICE) database and validate our findings.
\vspace{0.1cm}

\end{abstract}

\section{Introduction}
Fuzzy Sketches have been introduced to handle differences occurring
between 
two captures of biometric data, viewed as errors over a codeword. Many papers give applications of this technique for cryptographic purposes \cite{BoyenDKOS05,CoZe04a,CoZe04b,DavidaF99,DodisRS04,JuelsW99,TuylsG04}  but only a few investigate what are the best codes for this decoding problem, e.g. \cite{HaoAD06}, and how to find them. 
This issue is addressed here.

\subsection{Biometric matching and errors correction}
Typically, a biometric-based recognition scheme consists of two phases: The enrollment phase where a biometric template $b$ is measured from a 
user $U$ and then registered in a token or a database. The second phase -- the verification -- captures a new biometric sample $b'$ from $U$ 
and compares it to the reference data via a matching function. According to some underlying measure $\mu$ and some recognition threshold 
$\tau$, $b'$ will be accepted as a biometric measure of $U$ if $\mu(b,b')\leq \tau$, else rejected. 
Mainly two kinds of errors are associated to this scheme: 
False Reject (\highlight {FR}), when a matching user, i.e. a legitimate user, is rejected; 
False Acceptance (\highlight {FA}), when a non-matching one, e.g. an impostor, is accepted. 

Note that, when the threshold increases, the \highlight {FR}'s rate 
(\highlight {FRR}) decreases while the 
\highlight {FA}'s rate (\highlight {FAR}) grows, and conversely.

Our methods will resort to information theory and coding. For more background, notation and classical results, the reader is refered to \cite{CoverThomas} and \cite{McWSloane} in these two fields respectively.

Assuming that the templates live in the Hamming space 
$\mathcal{H}=\{0,1\}^n$ equipped with the Hamming distance $d_\mathcal{H}$, 
the main idea of fuzzy sketches, as introduced in \cite{JuelsW99}, is
to convert the matching step into an error-correcting one. 
Let $C$ be an error-correcting code included in $\mathcal{H}$: 
\begin{itemize}
\item During the enrollment phase, 
      one stores $z= c\oplus b$, where $c$ is a random codeword in $C$,
\item During the verification phase, 
one tries to correct the corrupted codeword $z\oplus
b'=c\oplus(b\oplus b')$. Note that when the Hamming distance 
$d_\mathcal{H}(b,b')$ is small, recovering $c$ from $c\oplus (b\oplus
b')$ is, in principle, possible.
\end{itemize}
The correction capacity of $C$ may thus be equal to $\tau$ if we do not want to alter the \highlight {FRR} and the \highlight {FAR} of the system. Unfortunately, the difference between two measures of one biometric source can be very important, whereas the correction capacity of a code 
is structurally constrained. Moreover, the size of the code should not be too small, to prevent $z$ from 
revealing too much information about the template $b$.

\subsection{Organization of this work}
In a first part, we look for theoretical limits. 
We first modelize our problem with a binary erasure-and-error channel. 
Given a database of biometric data, we then give a method for finding an upper bound on the underlying error correction capacity. 

In a second part, restricting ourselves to iris biometric data and illustrating our method with iterative min-sum decoding of product codes, 
we provide parameters that put  our performances close to the 
theoretical limit.

\section{Model}
We consider two separate channels with a noise model based on  the
differences between any two biometric templates. 
\begin{itemize}
\item The first channel, called the \highlight {matching channel}, is generated by errors $b\oplus b'$ where $b$ and $b'$ come from the same user $U$.
\item The second channel, the \highlight {non-matching channel}, is generated by errors where $b$ and $b'$ come from different biometric sources.
\end{itemize} 
In a practical biometric system, the number of errors in the \highlight {matching channel} is on average lower than in the \highlight {non-matching channel}.

Moreover, the templates are not restricted to a constant length. Indeed, when a sensor captures biometric data, we want to keep the maximum quantity of information but it is rarely possible to capture the same amount of data twice -- for instance an iris may be occulted by eyelids -- hence the templates are of variable length. This variability can be smoothed by forming a list of erasures, i.e. the list of coordinates where 
they occur. 
More precisely, in coding theory, an erasure in the received message 
is an unknown symbol at a known location. 
We thus have an erasure-and-error decoding problem on the 
\highlight {matching channel}. 
Simultaneously, to keep the \highlight {FAR} low, we want a decoding
success to be unlikely on the \highlight {non-matching channel}~: to
this end we impose bounds on the correction capacity.

In the sequel, we deal with binary templates with at most $N$ bits 
and assume, for the theoretical analysis that follows, that the
probabilities of error and erasure on each bit are independent.
Note that resorting to interleaving makes this hypothesis valid for
all practical purposes.

\subsection{Theoretical limit}\label{sec:theoretic}
Our goal is to estimate the capacity, in the Shannon sense \cite{Sh48}, of the matching channel when we work with a code of a given dimension. Namely, we want to know the maximum number of errors and erasures between two biometric measures that we can manage with fuzzy sketches for this code. 

Starting with a representative range of matching biometric data, the theorem below gives an easy way to estimate the lowest achievable 
\highlight {FRR}. The idea is to check whether the best possible code 
with the best generic decoding algorithm, 
i.e. a \highlight {maximum-likelihood} (\highlight {ML})
 decoding algorithm 
(which systematically outputs the most likely codeword), 
would succeed in correcting the errors. 

\begin{thm}\label{thm}
Let $k\in\N^*$, $C$ be a binary code of length $N$ and size $2^k$, and $m$ a random received message, from a random codeword of $C$, of length $N$ with $w_{n}$ errors and $w_{e}$ erasures. Assume that $C$ is an optimal code with respect to $N$ and $k$, equipped with an \highlight {ML} decoder. 

If $\frac{w_n}{N-w_e} > \theta$ then $m$ is only decodable with a 
negligible probability, where $\theta$ is such that the Hamming sphere of radius $(N-w_e)\theta$ in $\F_2^{N-w_e}$ contains $2^{N-w_e-k}$ elements.
\end{thm}
\begin{proof}
In the case of errors only (i.e. no erasures) with error-rate $p:=w_e/N$ , 
the canonical second theorem of Shannon asserts that there are families of codes 
with (transmission) rate $R:= k/n$ coming arbitrarily close to the channel 
capacity $\kappa(p)$, decodable with ML-decoding and a vanishing (in $N$) word error probability $P_e$.

In this case, $\kappa(p)= 1-h(p)$, where $h(p)$ is the (binary) entropy 
function ($\log$'s are to the base 2): 
$$h(x)= -x \log x -(1-x) \log (1-x).$$

Furthermore, $P_e$ displays a threshold phenomenon: for any rate arbitrarily close to, but above capacity and any family
of codes, $P_e$ tends to 1 when $N$ grows.

Equivalently, given $R$, there exists an error-rate threshold of 
$$p=h^{-1} (1-R),$$
$h^{-1}$ being the inverse of the
entropy function.

Back to the errors-and-erasures setting now. Our problem is to decode to the codeword nearest to the received word 
on the {\it nonerased} positions.

Thus we are now faced with a punctured code with length $N-w_e$, size
$2^k$, transmission rate $R' := k/(N-w_e)$ and required to sustain an error-rate 
 $p':= \frac{w_n}{N-w_e}$. 

By the previous discussion, if
$$p'> \theta := h^{-1} (1-R'),$$
NO code and NO decoding procedure exist with a non-vanishing probability
of success.

To conclude the proof, use the classical Stirling approximation for the size
of a Hamming sphere of radius $\alpha M$ in $\F_2^M$ by $2^{h(\alpha M)}$.
\end{proof}

Practical implications of this theorem are illustrated in Table \ref{table:limits}, Sec. \ref{sec:results}.

\section{Application}
\subsection{Description of the two-dimensional iterative min-sum decoding algorithm}\label{sec:minsum}

A binary {\it linear}  error-correcting code $C$ is a vector subspace
of $\F_2^N$. 
The minimum distance $d_{min}$ of $C$ is the smallest Hamming distance between two distinct codewords. 
When $k$ is the dimension of the subspace $C$, i.e. when it contains $2^k$ codewords, $C$ is denoted by $[N,k,d_{min}]_2$.
 The correction capacity $t$ of $C$ is the radius of the largest Hamming ball for which, for any $x\in\F_2^N$, there is at most one codeword in the ball of radius $t$ centered on $x$. 
Clearly,  $t=\lfloor (d_{min}-1)/2\rfloor$. An altered codeword with $w_n$ errors and $w_e$ erasures can always be corrected 
(by ML decoding) provided $2w_n+w_e<d_{min}$. 
However, if the code admits an iterative decoding algorithm, practical results overtake this limitation.

We will work with product codes together with a specific iterative decoding algorithm described below.  
A product code $C=C_1\otimes C_2$ is constructed from two codes: 
$C_1[N_1,k_1,d_1]_2$ and $C_2[N_2,k_2,d_2]_2$. 
The codewords of $C$ can be viewed as matrices of size $N_2\times N_1$ whose rows are codewords of $C_1$ and columns are codewords of $C_2$. 
This yields a $[N_1\times N_2, k_1\times k_2, d_1\times d_2]$ code. 
When $k_1$ and $k_2$ are small enough for $C_1$ and $C_2$ to be
decoded exhaustively
a very efficient iterative decoding algorithm is available, namely
the {\em min-sum} decoding algorithm. Min-sum decoding of LDPC codes
was developed by Wiberg \cite{wiberg} as a particular instance of
message passing algorithms. In a somewhat different setting it was
also proposed by Tanner \cite{tanner}
for decoding generalized LDPC (Tanner) codes. The variant we will
be using is close to Tanner's algorithm and is adapted to product
codes. Min-sum is usually considered to perform slightly worse
than the more classical sum-product message passing algorithm on
the Gaussian, or binary-symmetric channels, but it is specially
adapted to our case where knowledge of the channel is poor, and the
emphasis is simply to use the Hamming distance as the appropriate
basic cost function. 

Let $(x_{ij})$ be a vector of $\{0,1\}^{N_1\times N_2}$.
The min-sum algorithm associates to every coordinate $x_{ij}$ a
cost function $\kappa_{ij}$ for every iteration of the algorithm.
The cost functions are defined on the set $\{0,1\}$.
The initial cost function $\kappa_{ij}^0$ is defined by
$\kappa_{ij}^0(x) = 0$ if the received symbol on coordinate $(ij)$ is
$x$ and $\kappa_{ij}^0(x) = 1$ if the received symbol is $1-x$.

A {\em row} iteration of the algorithm takes an {\em input} cost function
$\kappa_{ij}^{in}$ and produces an {\em output} cost function
$\kappa_{ij}^{out}$. The algorithm first computes, for every row $i$
and for every codeword $c=(c_1\ldots c_{N_1})$ of $C_1$, the {\em sum}
$$\kappa_i(c) = \sum_{j=1}^{N_1}\kappa_{ij}^{in}(c_j)$$
which should be understood as the cost of putting codeword $c$ on row $i$.
The algorithm then computes, for every $i,j$, $\kappa_{ij}^{out}$ defined as the
following {\em min}, over the set of codewords of $C_1$,
$$\kappa_{ij}^{out}(x) = \min_{c\in C_1, c_j=x}\kappa_i(c).$$
This last quantity should be thought of as the minimum cost of putting the
symbol $x$ on coordinate $(ij)$ while satisfying the row constraint.

A {\em column} iteration of the algorithm is analogous to a row
iteration, with simply the roles of the row and column indexes reversed, and
code $C_2$ replacing code $C_1$. Precisely we have
  $$\kappa_j(c) = \sum_{i=1}^{N_2}\kappa_{ij}^{in}(c_i)$$
and
  $$\kappa_{ij}^{out}(x) = \min_{c\in C_2, c_i=x}\kappa_j(c).$$

\begin{figure}[hbtp!]
\beno
\\
i \left(
\begin{array}{ccc}
& \vdots & \\
\hline
\kappa_{i1}^{in} & \cdots & \kappa_{iN_1}^{in} \\
\hline
&  \vdots & \\
\end{array}
\right)
\eeno
$$\Downarrow\hspace{1cm}$$
$$\kappa_{ij}^{out}(x) = \min_{c\in C_1, c_j=x} \sum_{k=1}^{N_1}\kappa_{ik}^{in}(c_k)$$
$$\Downarrow\hspace{1cm}$$
\beno
i \left(
\begin{array}{ccc}
 & \vdots &  \\
\hline
\cdots & \kappa_{ij}^{out} & \cdots  \\
\hline
 & \vdots &   \\
\end{array}
\right)\hspace{0.4cm}
\eeno
$$
\begin{array}{c}
 \Downarrow\\[4mm]
 j
\end{array}\hspace{1cm}$$
\vspace{-2mm}
\beno
\left(
\begin{array}{c|c|c}
& \kappa_{1j}^{in} & \\
& \vdots & \\
\cdots & \vdots & \cdots \\
 & \vdots & \\
& \kappa_{N_2j}^{in}& \\
\end{array}
\right)\hspace{0.4cm}
\eeno
$$\Downarrow\hspace{1cm}$$
$$\kappa_{ij}^{out}(x) = \min_{c\in C_2, c_i=x} \sum_{l=1}^{N_2}\kappa_{lj}^{in}(c_l)$$
$$
\begin{array}{c}
 \Downarrow\\[4mm]
 j
\end{array}\hspace{1cm}$$
\vspace{-2mm}
\beno
\left(
\begin{array}{c|c|c}
& \vdots & \\
& \vdots & \\
\cdots & \kappa_{ij}^{out} & \cdots \\
 & \vdots & \\
& \vdots & \\
\end{array}
\right)\hspace{0.4cm}
\eeno
\caption{A row iteration followed by a column one}\label{tab:minsum}
\end{figure}
  
The algorithm alternates row and column iterations as illustrated by Fig. \ref{tab:minsum}. After a given
number of iterations (or before, if we find a codeword) it stops, and the value of every symbol
$x_{ij}$ is put at $x_{ij}=x$ if $\kappa_{ij}^{out}(x)<\kappa_{ij}^{out}(1-x)$.
If  $\kappa_{ij}^{out}(x)= \kappa_{ij}^{out}(1-x)$ then the value of
$x_{ij}$ stays undecided (or erased).

The following theorem is fairly straightforward to prove and illustrates the
power of min-sum decoding.

\begin{thm}
  If the number of errors is less than $d_1d_2/2$, then two iterations 
  of min-sum decoding of the product code $C_1\otimes C_2$ recover
  the correct codeword.
\end{thm}

\subsection{Our setting}
To validate our approach, we now present the results of experiments on
a practical iris database where we obtain correction performances close 
to the theoretical limit. 

The database used for these experiments is the ICE (Phase I) database \cite{ICE0,ICE} which contained 2953 images from 244 different eyes. A 256-byte (2048 bits) iris template, together with a 256-byte mask, is computed from each iris image using the algorithm reported in \cite{Dau03}; the mask filters out the unreliable bits, i.e. stores the erasures indices of the iris template. 
The database is taken without any modification but two slight corrections: one eye is suppressed due to a very low quality and the side of another eye has been switched from left to right. Hence we keep 2952 images. Note that in the database, the number of images provided for each eye is variable: so the number of intra-eye matching verifications between two iris codes from the same eye is not constant. 
The same holds for the inter-eye matching between two iris codes from different eyes. Among all the combinations, its gives a set of 29827 intra-eye matching and about 4 million of inter-eye matching to check.

\begin{figure}[!htbp]
\centering
\includegraphics[width=8cm]{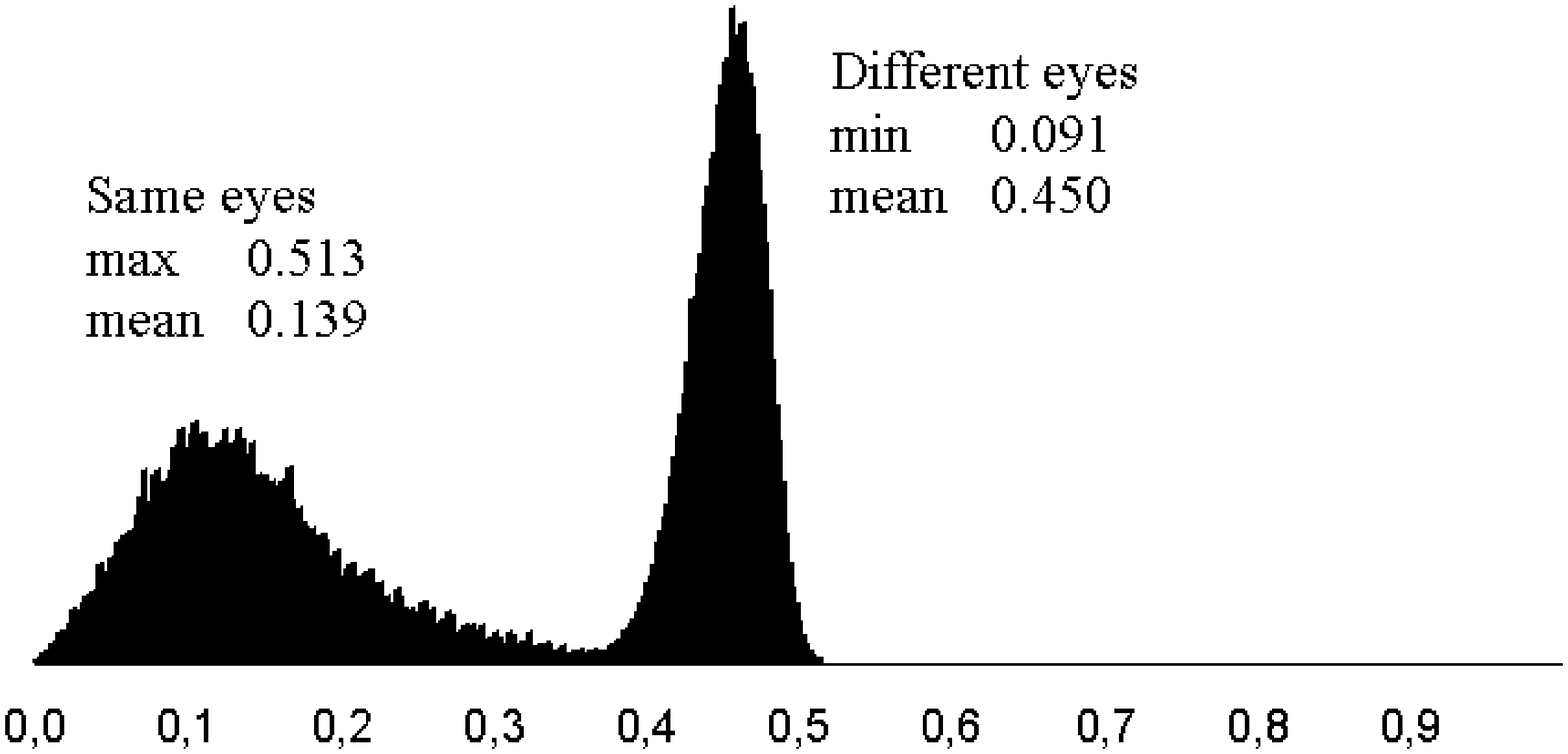}
\caption{Inter-eyes and intra-eye distributions}
\label{fig}
\end{figure}

The classical way to compare two iris codes $I_1, I_2$ with masks
$M_1, M_2$ is to compute the relative Hamming distance
$$\frac{||(I_1\oplus I_2)\cap M_1\cap M_2||}{||M_1\cap M_2||}$$ for
some rotations of the second template -- to deal with the iris
orientation's variation -- and to keep the lowest score. 
It gives the following distributions of matching scores (cf. Fig \ref{fig}) 
where we see an overlap between the two curves.
We also see that the number of errors to handle in the matching
channel is large (for instance at least $29\%$ of errors for a
\highlight {FRR} lower than $5\%$). On this channel, 
an additional difficulty originates from the number of erasures which
varies from 512 to 1977.

\subsection{Results on ICE database}\label{sec:results}
We have experimented with the algorithm described in section~\ref{sec:minsum} 
on this database with a particular choice for the code. In fact, the product code is constructed to fit with an array of 2048 bits, by using Reed-Muller codes \cite{Mu54,Re54} of order 1 which are known to have good weight distributions. A binary Reed-Muller code of order 1 in $m$ variables, abbreviated as $RM(1,m)$,  is a $[2^m,m+1,2^{m-1}]_2$ code. 
We chose to combine the $RM(1,6)$ with the $RM(1,5)$, leading to a product code of dimension 42 and codewords of length $64\times 32$.

As the density of errors and erasures in an iris code can be very high in some regions, we also added a randomly chosen interleaver to break 
this structure and increase the efficiency of the decoding
algorithm. In
so doing, we succeeded in obtaining a \highlight {FRR} of about $5.62\%$ 
for a very small \highlight {FAR} (lower than $10^{-5}$). This is in fact very close to the \highlight {FAR} obtained in a classical matching configuration for a similar \highlight {FRR}.

The overall size of the code could appear small from a cryptographic point of view, but following the theoretical analysis of 
section \ref{sec:theoretic}, it is difficult to expect much more while achieving a low \highlight {FRR} on this database. 
Indeed, from the distribution of errors and erasures on the \highlight {matching channel}, we obtain by Theorem \ref{thm} the practical limits 
which are reported in Table \ref{table:limits}.

\begin{table}[!hbtp]
\caption{Theoretical limits on ICE database}\label{table:limits}
\begin{center}
\begin{tabular}{|c|c|}
\hline
Code's dimension & Best theoretical FRR \\
\hline
\hline
42 & $2.49\%$ \\
\hline
64 & $3.76\%$ \\
\hline
80 & $4.87\%$ \\
\hline
128 & $9.10\%$ \\
\hline
\end{tabular}
\end{center}
\end{table}

{\bf Remark.}
In \cite{HaoAD06}, the fuzzy sketch scheme is applied with a
concatenated error-correcting code combining a Hadamard code and a
Reed-Solomon code. More precisely, the authors use a Reed-Solomon code
of length 32 over $\F_{2^7}$ (with a correction capacity $t_{RS}<16$) and a Hadamard code of order 6 and length 64 (with a correction capacity $t_H=15$): a codeword of 2048 bits is in fact constructed as a set of 32 blocks of 64 bits where each block is a codeword of the underlying Hadamard code. As explained in \cite{HaoAD06}, the Hadamard code is introduced to deal with the background errors and the Reed-Solomon code to deal with the bursts (e.g. caused by eyelashes, reflections, $\ldots$).

Note that in this scheme, the model is not exactly the same as ours, as the masks are not taken into account. 
Moreover, the quality of the database used in \cite{HaoAD06} is better than for the ICE database. 
Actually, \cite{HaoAD06} reports very good results on their experiments with a database of 700 images, but 
the codes do not seem appropriate to our case as our experiment on the
ICE 
database gave a too large rate of \highlight {FR} 
(e.g. $10\%$ of \highlight {FR} with $0.80\%$ of \highlight {FA}), even for the smallest possible dimension of the Reed-Solomon code when $t_{RS}=15$.

\section{Conclusion}

We derived explicit upper bounds on the correction capacity of Fuzzy Sketches on iris-based biometrics.
We then showed how the two-dimensional iterative min-sum decoding algorithm achieves correction performance close to the optimal decoding 
rate. Our results were validated on a typical iris database.

\end{document}